\documentclass[amsmath,amssymb]{revtex4}
\usepackage{graphicx}
\usepackage{amscd,amsmath,amsthm,amsfonts,amssymb}

\def\PT{$\cal{PT}$}
\def\x{\mathbf{x}}
\def\[{\begin{equation}}
\def\]{\end{equation}}

\advance\textheight -16mm

\numberwithin{equation}{section}

\begin{document}
\title{Transformations between nonlocal and local integrable equations}
\author{Bo Yang$^{1,2}$ and Jianke Yang$^{2*}$}
\address{
{\small\it $^1$Shanghai Key Laboratory of Trustworthy Computing, East China Normal University, Shanghai 200062, China} \\
{\small\it $^2$Department of Mathematics and Statistics, University of Vermont, Burlington, VT 05401, U.S.A} \\
{\normalsize \small \it * Corresponding author, email address: jyang@math.uvm.edu}}

\begin{abstract}
Recently, a number of nonlocal integrable equations, such as the \PT-symmetric nonlinear Schr\"odinger (NLS) equation and \PT-symmetric Davey-Stewartson equations, were proposed and studied. Here we show that many of such nonlocal integrable equations can be converted to local integrable equations through simple variable transformations. Examples include these nonlocal NLS and Davey-Stewartson equations, a nonlocal derivative NLS equation, the reverse space-time complex modified Korteweg-de Vries (CMKdV) equation, and many others. These transformations not only establish immediately the integrability of these nonlocal equations, but also allow us to construct their analytical solutions from solutions of the local equations. These transformations can also be used to derive new nonlocal integrable equations. As applications of these transformations, we use them to derive rogue wave solutions for the partially \PT-symmetric Davey-Stewartson equations and the nonlocal derivative NLS equation. In addition, we use them to derive multi-soliton and quasi-periodic solutions in the reverse space-time CMKdV equation. Furthermore, we use them to construct many new nonlocal integrable equations such as nonlocal short pulse equations, nonlocal nonlinear diffusion equations, and nonlocal Sasa-Satsuma equations.
\end{abstract}


\maketitle

\section{Introduction}
The study of integrable nonlinear wave equations has a long history \cite{Ablowitz1981,Zakharov1984,Faddeev1987,Ablowitz1991,Yang2010}. Most of those integrable equations are local equations, i.e., the solution's evolution depends only on the local solution value and its local space and time derivatives. The Korteweg-de Vries equation and the nonlinear Schr\"odinger (NLS) equation are such examples.

Recently, a number of new nonlocal integrable equations were proposed and studied \cite{AblowitzMussPRL2013,AblowitzMussPRE2014,Yan,Khara2015,AblowitzMussNonli2016,Zhu1,Fokas2016,Lou,Lou2,AblowitzMussSAPM,Chow,ZhoudNLS,ZhouDS,HePTDS,HePPTDS,Zhu2,Zhu3,Gerdjikov2017,
Ablowitz_arxiv}. The first such nonlocal equation was the \PT-symmetric NLS equation \cite{AblowitzMussPRL2013}
\[ \label{e:PTNLS}
iu_t(x,t)+u_{xx}(x,t)+2\sigma u^2(x,t)u^*(-x,t)=0,
\]
where $\sigma=\pm 1$ is the sign of nonlinearity (with the plus sign being the focusing case and minus sign the defocusing case), and the asterisk * represents complex conjugation. Notice that here, the solution's evolution at location $x$ depends on not only the local solution at $x$, but also the nonlocal solution at the distant position $-x$. That is, solution states at distant locations $x$ and $-x$ are directly coupled, reminiscent of quantum entanglement between pairs of particles. Eq. (\ref{e:PTNLS}) was called \PT-symmetric because the nonlinearity-induced potential $V\equiv \sigma u(x,t)u^*(-x, t)$ satisfies the \PT symmetry condition $V^*(x,t)=V(-x,t)$. In addition, this equation is invariant under the action of the \PT operator, i.e., the joint transformations $x\to -x$, $t\to -t$ and complex conjugation [hence if $u(x,t)$ is a solution, so is $u^*(-x,-t)$)]. It is noted that \PT symmetric systems, which behave like conservative systems despite gain and loss, attracted a lot of attention in optics and other physical fields in recent years \cite{Kivsharreview,Yangreview}. The application of this \PT-symmetric NLS equation for an unconventional system of magnetics was reported in \cite{PTNLSmagnetics}.

Following this nonlocal \PT-symmetric NLS equation, other new nonlocal integrable equations were quickly reported. Examples include the fully \PT-symmetric and partially \PT-symmetric Davey-Stewartson (DS) equations
\cite{Fokas2016,AblowitzMussSAPM}, the nonlocal derivative NLS equation \cite{ZhoudNLS}, the reverse space-time complex modified Korteweg-de Vries (CMKdV) equation \cite{AblowitzMussNonli2016,Zhu3}, the reverse time NLS equation \cite{AblowitzMussSAPM}, the reverse space-time NLS equation \cite{Lou,AblowitzMussSAPM}, and many others. These nonlocal equations are distinctly different from local equations for their novel space and/or time coupling, which could induce new physical effects and thus inspire novel physical applications. Indeed, solution properties in some of these nonlocal equations have been analyzed by the inverse scattering transform method, Darboux transformation or the Hirota bilinear method, and interesting behaviors such as finite-time solution blowup \cite{AblowitzMussPRL2013} and the simultaneous existence of soliton and kink solutions \cite{Zhu2} have been revealed.

In this article, we report that many of these nonlocal integrable equations can be converted to their local integrable counterparts through simple variable transformations. Such nonlocal equations include the \PT-symmetric NLS and DS equations, the nonlocal derivative NLS equation, the reverse space-time CMKdV equation, and many others. This conversion puts these nonlocal equations in a totally different light and opens up a totally new way to study their solution behaviors. First of all, this conversion
immediately establishes the integrability of these nonlocal equations. Secondly, it allows us to construct analytical solutions of these nonlocal equations from solutions of their local counterparts. Thirdly, it can be used to derive new nonlocal integrable equations. As applications of these transformations, we use them to derive rogue wave solutions for the partially \PT-symmetric DS equations and the nonlocal derivative NLS equation. In addition, we use them to derive multi-soliton and quasi-periodic solutions in the reverse space-time CMKdV equation. Furthermore, we use them to construct many new nonlocal integrable equations such as nonlocal short pulse equations, nonlocal nonlinear diffusion equations, nonlocal Sasa-Satsuma equations and nonlocal Chen-Lee-Liu equations.

\section{Transformations between nonlocal and local integrable equations}
In this section, we present transformations which convert many nonlocal integrable equations to their local counterparts.

Our first example is the \PT-symmetric NLS equation (\ref{e:PTNLS}). Under the variable transformations
\[ \label{transNLS}
x=i\hat{x}, \hspace{0.2cm} t=-\hat{t}, \hspace{0.2cm} u(x,t)=\hat{u}(\hat{x},\hat{t}),
\]
this nonlocal equation becomes
\[ \label{e:PTNLSL}
i\hat{u}_{\hat{t}}(\hat{x},\hat{t})+\hat{u}_{\hat{x}\hat{x}}(\hat{x},\hat{t})-2\sigma \hat{u}^2(\hat{x},\hat{t})\hat{u}^*(\hat{x},\hat{t})=0,
\]
which is the local NLS equation but with the opposite sign of nonlinearity. In other words, the \PT-symmetric focusing NLS equation is converted to the local defocusing NLS equation, and the \PT-symmetric defocusing NLS equation is converted to the local focusing NLS equation.
The key reason for this nonlocal to local conversion is that, in the nonlocal equation (\ref{e:PTNLS}), $x$ is treated real when taking the complex conjugate $u^*(-x,t)$. But under the $x=i\hat{x}$ transformation with real $\hat{x}$, $x$ becomes imaginary. In this case, when taking the complex conjugate of $u(-x,t)$, the sign of $x$ flips, turning the nonlocal term $u^*(-x,t)$ in (\ref{e:PTNLS}) to the local term $\hat{u}^*(\hat{x},\hat{t})$ in (\ref{e:PTNLSL}).

Following similar ideas, we can transform many more nonlocal integrable equations to their local counterparts. Some examples are listed below.

\begin{enumerate}

\item Consider the \PT-symmetric DS equations \cite{Fokas2016,AblowitzMussSAPM}
\begin{eqnarray}  \label{PTDS}
&& \hspace{-1.4cm} iA_t(x,y,t) = A_{xx}(x,y,t)+\sigma^2 A_{yy}(x,y,t) +\left[\epsilon A(x,y,t)A^*(-x,-y,t)-2Q(x,y,t)\right]A(x,y,t),  \nonumber \\
&& \hspace{-1.4cm} Q_{xx}(x,y,t)-\sigma^2Q_{yy}(x,y,t)=\epsilon \left[A(x,y,t)A^*(-x,-y,t)\right]_{xx},
\end{eqnarray}
where $\sigma^2=\pm 1$ is the equation-type parameter (with $\sigma^2=1$ being DSI and $\sigma^2=-1$ DSII), and $\epsilon=\pm 1$ is the sign of nonlinearity. Under the variable transformations
\[
x=i\hat{x}, \hspace{0.2cm} y=i\hat{y}, \hspace{0.2cm} t=-\hat{t}, \hspace{0.2cm}
A(x,y,t)=\widehat{A}(\hat{x},\hat{y},\hat{t}), \hspace{0.2cm}  Q(x,y,t)=-\widehat{Q}(\hat{x},\hat{y},\hat{t}),
\]
and dropping the bars, i.e., with
\[x\to ix, \hspace{0.2cm} y\to iy, \hspace{0.2cm} t\to -t, \hspace{0.2cm} Q\to -Q,\]
these \PT-symmetric DS equations are
converted to the following local (classical) DS equations
\begin{eqnarray}
&& \hspace{-1.4cm} iA_{t}(x,y,t)=A_{xx}(x,y,t)+\sigma^2 A_{yy}(x,y,t) +\left[-\epsilon A(x,y,t)A^*(x,y,t)-2Q(x,y,t)\right]
A(x,y,t), \nonumber \\
&& \hspace{-1.4cm} Q_{xx}(x,y,t)-\sigma^2Q_{yy}(x,y,t)=-\epsilon \left[A(x,y,t)A^*(x,y,t)\right]_{xx}.
\end{eqnarray}
Similar to the \PT-symmetric NLS equation above, the sign of nonlinearity $\epsilon$ has switched after the nonlocal-to-local conversion, but the sign of $\sigma^2$ remains the same. Thus, the \PT-symmetric focusing DSI equations are converted to local defocusing DSI equations, and so on.

\item Consider the partially \PT-symmetric DS equations \cite{Fokas2016,AblowitzMussSAPM}
\begin{eqnarray} \label{PPTDS}
&& \hspace{-1.4cm} iA_t(x,y,t)=A_{xx}(x,y,t)+\sigma^2 A_{yy}(x,y,t) +\left[\epsilon A(x,y,t)A^*(-x,y,t)-2Q(x,y,t)\right]A(x,y,t), \nonumber \\
&& \hspace{-1.4cm} Q_{xx}(x,y,t)-\sigma^2Q_{yy}(x,y,t)=\epsilon \left[A(x,y,t)A^*(-x,y,t)\right]_{xx},
\end{eqnarray}
where $\sigma^2=\pm 1$ and $\epsilon=\pm 1$. Under the variable transformations
\[ \label{transPPTDS}
x\to ix, \hspace{0.2cm} t\to -t, \hspace{0.2cm} Q\to -Q,
\]
these nonlocal DS equations reduce to the local DS equations
\begin{eqnarray} \label{PPTDSL}
&& \hspace{-1.4cm} iA_{t}(x,y,t)=A_{xx}(x,y,t)-\sigma^2 A_{yy}(x,y,t) +\left[-\epsilon A(x,y,t)A^*(x,y,t)-2Q(x,y,t)\right]
A(x,y,t), \nonumber \\
&& \hspace{-1.4cm} Q_{xx}(x,y,t)+\sigma^2Q_{yy}(x,y,t)=-\epsilon \left[A(x,y,t)A^*(x,y,t)\right]_{xx}.
\end{eqnarray}
Here, after the nonlocal-to-local conversion, not only the sign of nonlinearity $\epsilon$, but also the equation-type parameter $\sigma^2$, switches. Thus, the partially \PT-symmetric focusing DSI equations are converted to local defocusing DSII equations, and so on.

The equations (\ref{PPTDS}) are partially \PT-symmetric in $x$. Similar equations partially \PT-symmetric in $y$ can also be converted to their local counterparts through the sole transformation $y\to iy$. Under this conversion, the sign of $\sigma^2$ switches, but not the sign of $\epsilon$.

\item Consider the nonlocal derivative NLS equation \cite{ZhoudNLS}
\[ \label{e:PTDNLS}
iu_t(x,t)+u_{xx}(x,t)+\sigma \left[u^2(x,t)u^*(-x,t)\right]_x=0,
\]
where $\sigma=\pm 1$. Under the variable transformations
\[ \label{transDNLS}
x\to ix, \hspace{0.15cm} t\to -t,
\]
this nonlocal equation becomes the local derivative NLS equation \cite{KaupNewell}
\[
\label{e:PTDNLSL}
iu_t(x,t)+u_{xx}(x,t)+i\sigma \left[u^2(x,t)u^*(x,t)\right]_x=0.
\]

\item Consider the reverse space-time CMKdV equation \cite{AblowitzMussNonli2016,Zhu3}
\[ \label{e:NCMKdV}
q_{t}(x,t)+q_{xxx}(x,t)+6\sigma q(x,t)q^{*}(-x,-t)q_{x}(x,t)=0,
\]
where $\sigma=\pm 1$.
Under the variable transformations
\[ \label{transCMKdV}
x\rightarrow ix,\ t\rightarrow -it,
\]
this equation become the following local (classical) CMKdV equation
\[ \label{e:CMKdV}
q_{t}(x,t)+q_{xxx}(x,t)-6\sigma q(x,t)q^{*}(x,t)q_{x}(x,t)=0.
\]
Note that the sign of nonlinearity has flipped under this conversion.

\item Consider the multidimensional reverse space-time nonlocal three wave interaction equations \cite{AblowitzMussSAPM}
\begin{eqnarray*}
&& \hspace{-1.2cm} Q_{1,t}(\x,t)+\mathbf{C}_1\cdot \nabla Q_1(\x,t)=\sigma_1 Q_2^*(-\x,-t)Q_3^*(-\x,-t), \nonumber \\
&& \hspace{-1.2cm} Q_{2,t}(\x,t)+\mathbf{C}_2\cdot \nabla Q_2(\x,t)=\sigma_2 Q_1^*(-\x,-t)Q_3^*(-\x,-t), \\
&& \hspace{-1.2cm} Q_{3,t}(\x,t)+\mathbf{C}_3\cdot \nabla Q_3(\x,t)=\sigma_3 Q_1^*(-\x,-t)Q_2^*(-\x,-t), \nonumber
\end{eqnarray*}
where $\sigma_j=\pm 1$, $j=1, 2, 3$, $\sigma_1\sigma_2\sigma_3=-1$, $\x$ is a multidimensional spacial variable, and $\mathbf{C}_1$, $\mathbf{C}_2$, $\mathbf{C}_3$ are constant vectors. Under the variable transformations
\[
\x\to i\hspace{0.04cm} \x, \quad t\to i \hspace{0.04cm} t,  \nonumber
\]
the above nonlocal three wave interaction equations reduce to the local counterparts~\cite{Ablowitz1981}
\begin{eqnarray*}
&& \hspace{-0.8cm} Q_{1,t}(\x,t)+\mathbf{C}_1\cdot \nabla Q_1(\x,t)=i\sigma_1 Q_2^*(\x,t)Q_3^*(\x,t), \nonumber \\
&& \hspace{-0.8cm} Q_{2,t}(\x,t)+\mathbf{C}_2\cdot \nabla Q_2(\x,t)=i\sigma_2 Q_1^*(\x,t)Q_3^*(\x,t), \\
&& \hspace{-0.8cm} Q_{3,t}(\x,t)+\mathbf{C}_3\cdot \nabla Q_3(\x,t)=i\sigma_3 Q_1^*(\x,t)Q_2^*(\x,t). \nonumber
\end{eqnarray*}

\end{enumerate}
In addition to the above nonlocal integrable equations, many others, such as the vector or matrix extensions of
the \PT-symmetric NLS equations \cite{Yan,Zhu1,AblowitzMussSAPM}, can also be converted to local integrable equations through similar transformations.

These transformations between nonlocal and local integrable equations offer a totally different way of studying these nonlocal equations, and they can be used for many purposes. First of all, these transformations immediately establish the integrability of the underlying nonlocal equations in view of the integrability of their local counterparts. Secondly, these transformations allow us to obtain the infinite number of conservation laws for the nonlocal equations from those of local equations. For example, from the first two conserved quantities of the NLS equation (\ref{e:PTNLSL}),
\[
I_1=\int_{-\infty}^\infty \hat{u}(\hat{x},\hat{t})\hat{u}^*(\hat{x},\hat{t}) d\hat{x}, \quad I_2=\int_{-\infty}^\infty \hat{u}^*(\hat{x},\hat{t})\hat{u}_{\hat{x}}(\hat{x},\hat{t}) d\hat{x},  \nonumber
\]
we immediately obtain through variable transformations (\ref{transNLS}) the first two conserved quantities of the \PT-symmetric NLS equation (\ref{e:PTNLS}) \cite{AblowitzMussPRL2013,AblowitzMussNonli2016}
\[
I_1=\int_{-\infty}^\infty u(x,t)u^*(-x, t)dx, \quad I_2=\int_{-\infty}^\infty u^*(-x, t)u_x(x,t) dx.   \nonumber
\]
Thirdly, these transformations allow us to construct analytical solutions of nonlocal equations from those of local ones. Examples of this will be presented in the next three sections. Fourthly, these transformations can be used to derive new nonlocal integrable equations from their local counterparts. This will be demonstrated in Sec. \ref{sec:newnonlocal}.

It is noted that the solution construction of nonlocal equations through these transformations may not be as trivial as it seems. The reason is that well-behaved solutions of the local equations may become ill-behaved under these transformations. For instance, the soliton solution of the local focusing NLS equation (\ref{e:PTNLSL}) (with $\sigma=-1$),
\[
\hat{u}(\hat{x},\hat{t})=e^{i\hspace{0.04cm} \hat{t}} \mbox{sech}\hspace{0.03cm} \hat{x},
\]
under transformations (\ref{transNLS}), becomes a singular solution
\[
u(x,t)=e^{-i\hspace{0.04cm} t} \mbox{sec}\hspace{0.03cm} x
\]
of the nonlocal defocusing NLS equation (\ref{e:PTNLS}). Thus, in order to derive nonsingular solutions of nonlocal integrable equations through these transformations, one needs to choose solutions of local equations carefully.

\section{Rogue waves in partially \PT-symmetric DS equations}
In this section, we derive rogue wave solutions in partially \PT-symmetric DS equations (\ref{PPTDS}) using the transformation method. These rogue wave solutions have not been reported before to our best knowledge.

\subsection{Partially \PT-symmetric DSI equations}
Rogue waves are rational solutions. According to the transformations (\ref{transPPTDS}), rational solutions in partially \PT-symmetric DSI equations (\ref{PPTDS}) (with $\sigma^2=1$) can be obtained from rational solutions in local DSII equations (\ref{PPTDSL}). These latter solutions have been reported in \cite{Satsuma_Ablowitz,YangDSII}. Utilizing those solutions and the reverse variable transformations of (\ref{transPPTDS}), i.e., $x\to -ix$, $t\to -t$ and $Q\to -Q$, and accounting for the sign switching of the nonlinearity parameter $\epsilon$, rational solutions in the partially \PT-symmetric DSI equations (\ref{PPTDS}) can be obtained. By imposing parameter conditions on these rational solutions, rogue wave solutions in these nonlocal DSI equations can then be derived.

First, we consider fundamental rational solutions in the partially \PT-symmetric DSI equations (\ref{PPTDS}). These solutions are deduced from the fundamental rational solutions (11)-(12) of the local DSII equations in Ref. \cite{YangDSII} under the reverse variable transformations as
\begin{equation}  \label{PPTAxyt}
A(x,y,t)=\sqrt{2}\left[1-\frac{2i(-ia_2x+b_2y-\omega_2t+\theta_2)+1}{f}\right],
\end{equation}
\begin{equation}  \label{PPTQxyt}
Q(x,y,t)=\epsilon-(2\ln f)_{xx},
\end{equation}
where
\begin{equation} \label{e:fDSI}
f=(-ia_1x+b_1y-\omega_1t+\theta_1)^2+(-ia_2x+b_2y-\omega_2t+\theta_2)^2+\Delta,
\end{equation}
\begin{equation} \label{aa1a2}
a=a_1+ia_2, \hspace{0.2cm} b=b_1+ib_2, \hspace{0.2cm} \omega=\omega_1+i\omega_2, \hspace{0.2cm}
\theta=\theta_1+i\theta_2,
\end{equation}
\begin{equation} \label{abomega}
a\equiv \frac{p+\epsilon/p}{2}, \quad b\equiv \frac{p-\epsilon/p}{2}i,  \quad \omega\equiv \frac{p^2+1/p^2}{i}, \quad \Delta\equiv \frac{\epsilon|p|^2}{(|p|^2+\epsilon)^2},
\end{equation}
$p$, $\theta$ are free complex parameters, and $a_{1,2}, b_{1,2}, \omega_{1,2}, \theta_{1,2}$ are the real and imaginary parts of complex numbers $a, b, \omega, \theta$. Performing solution analysis analogous to that in \cite{YangDSII}, we find that this rational solution is a rogue wave when $p$ is purely imaginary. In this case, the solutions go to a constant background, $A\to \sqrt{2}$, $Q\to \epsilon$, as $t\to -\infty$. Since $p$ is imaginary, $a$ and $\omega$ are imaginary, and $b$ is real. Thus $a_1=b_2=\omega_1=0$, and the function $f$ in (\ref{e:fDSI}) becomes
\[
f=(b_1y+\theta_1)^2+(-ia_2x-\omega_2 t+\theta_2)^2+\Delta.
\]
This function is nonzero as long as $\omega_2t-\theta_2$ is nonzero. Thus, the above rogue wave is nonsingular as long as $\omega_2t-\theta_2\ne 0$. But when $\omega_2t-\theta_2=0$, i.e., at a critical time $t_c=\theta_2/\omega_2$, the function $f$ becomes zero at spatial positions where
\[ \label{e:hyperbola}
(b_1y+\theta_1)^2-a_2^2x^2+\Delta=0.
\]
In the generic case where $b_1\ne 0$ and $a_2\ne 0$, i.e., $p^2 \ne -1$, this equation defines a hyperbola on the $(x,y)$ plane.
Thus, this rogue wave, which arises from a constant background, develops finite time singularity on the entire hyperbola (\ref{e:hyperbola}) at the critical time $t_c=\theta_2/\omega_2$. In addition, this rogue wave exists for both signs of nonlinearity $\epsilon=\pm 1$. In the non-generic case where $p^2=-1$ and $\epsilon=1$, $a=0$, hence the above rogue wave is $x$-independent. Without $x$ dependence, the partially \PT-symmetric DSI equations (\ref{PPTDS}) degenerate to the local NLS equation with the spatial variable $y$, and the above rogue wave degenerates to the Peregrine rogue wave of the NLS equation \cite{Peregrine}. Note that the case of $p^2=-1$ and $\epsilon=-1$ is inadmissible since $\Delta$ is infinite here.

\begin{figure}[h!]
\centerline{\includegraphics[width=0.7\textwidth]{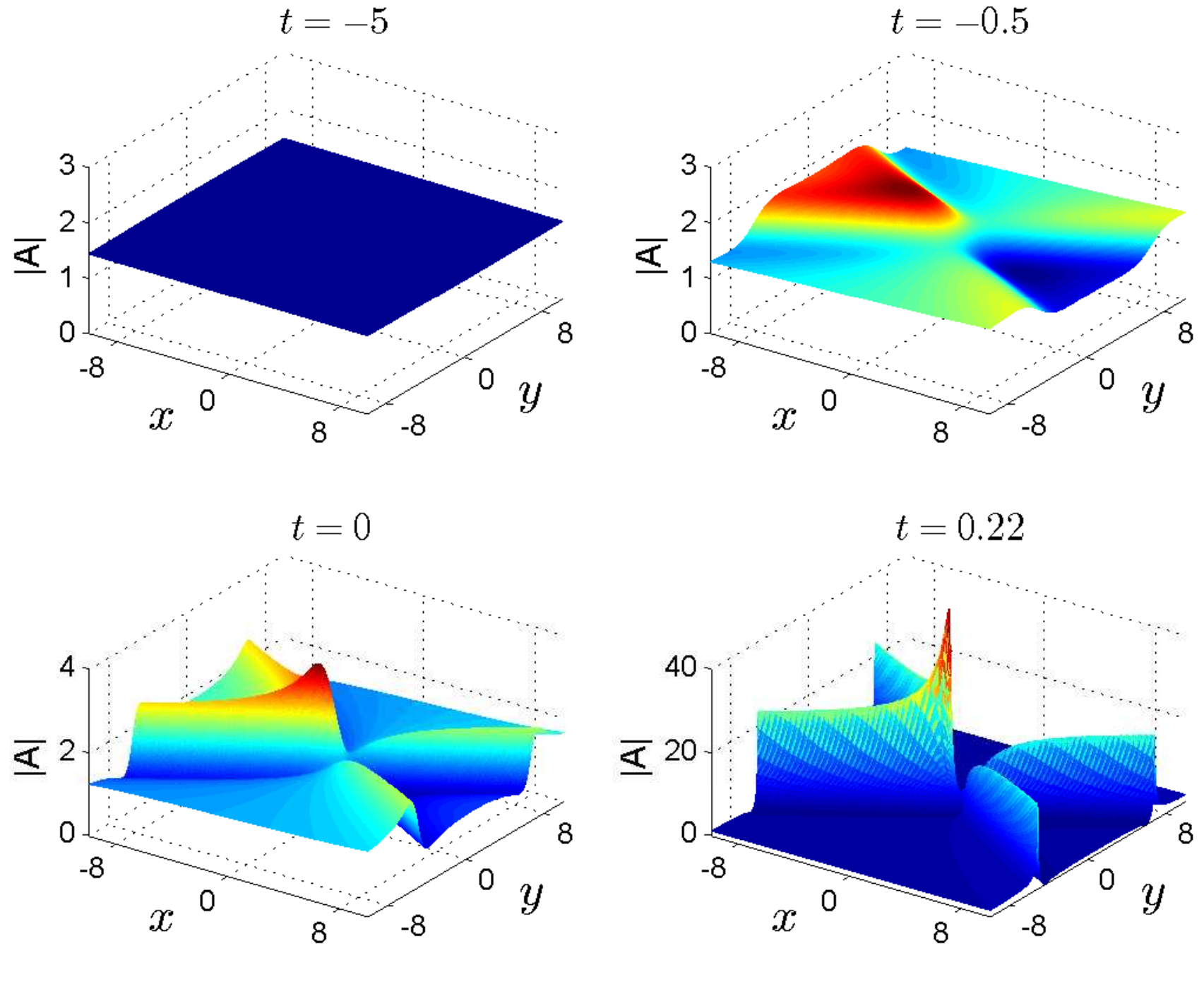}}
\caption{(color online) An exploding fundamental rogue wave (\ref{PPTAxyt}) in the partially \PT-symmetric DSI equations (\ref{PPTDS}) with $\sigma^2=1$, $\epsilon=1$, $p=0.5i$ and $\theta=1+i$.  }
\label{f:fig1}
\end{figure}

To illustrate this fundamental rogue wave, we choose $\epsilon=1$, $p=0.5i$ and $\theta=1+i$. The corresponding rogue wave is plotted in Fig.~\ref{f:fig1}. In this case, the finite-time singularity occurs at $t_c=4/17\approx 0.2353$, thus we only plotted solutions up to time $t=0.22$, shortly before the blowup.

 It is interesting to compare this fundamental rogue wave of the nonlocal DSI equation with that of the local DSII equation \cite{YangDSII}. First of all, the parameter conditions are very different. In the local DSII equation, rogue waves require $|p|=1$; if $|p|\ne 1$, the rational solution would be a two-dimensional lump moving on a constant background. In the nonlocal DSI equation (\ref{PPTDS}), rogue waves require $q$ to be purely imaginary. In this case, $|p|\ne 1$ generically, and thus the transformations (\ref{transPPTDS}) convert moving-lump solutions of the local DSII equation into rogue waves of the nonlocal DSI equation. Secondly, in the local DSII equation, rogue waves exist only when $\epsilon=-1$; but in the nonlocal DSI equation (\ref{PPTDS}), rogue waves exist for both signs of nonlinearity $\epsilon=\pm 1$. Thirdly, in the local DSII equation, fundamental rogue waves are line rogue waves; but in the nonlocal DSI equation, fundamental rogue waves have richer structures. Fourthly, in the local DSII equation, fundamental rogue waves never blow up in finite time; but in the nonlocal DSI equation, fundamental rogue waves generically blow up in finite time. Although some non-generic multi-rogue waves and higher-order rogue waves of the local DSII equation can also blow up in finite time, they only do so at a single spatial point, unlike the fundamental rogue waves of the nonlocal DSI equation where the blowup occurs on an entire hyperbola of the spatial plane.

Multi-rogue waves of the nonlocal DSI equations (\ref{PPTDS}) can be similarly derived from those of the local DSII equations in \cite{Satsuma_Ablowitz,YangDSII} under the reverse variable transformations $x\to -ix$, $t\to -t$, $Q\to -Q$ and the parameter conditions of $p_j \hspace{0.07cm} (1\le j\le n)$ being purely imaginary. These multi-rogue waves describe the nonlinear interaction of several individual fundamental rogue waves. Details are omitted.

It is noted that rogue waves in the fully \PT-symmetric DS equations (\ref{PTDS}) have been reported in \cite{HePTDS,HePPTDS}. Those rogue waves can also be derived using our transformation method in view of the conversion of these nonlocal equations to the local DS equations as discussed in the previous section.

\subsection{Partially \PT-symmetric DSII equations}
Rogue waves in partially \PT-symmetric DSII equations (\ref{PPTDS}), with $\sigma^2=-1$, can be obtained from the rational solutions in the local DSI equations (\ref{PPTDSL}) under the reverse of transformations (\ref{transPPTDS}). These rational solutions in the local DSI equations have been reported in \cite{Satsuma_Ablowitz,YangDSI}. Imposing suitable parameter restrictions, rogue waves in the nonlocal DSII equations (\ref{PPTDS}) will be obtained.

Fundamental rational solutions in the nonlocal DSII equations (\ref{PPTDS}) can be obtained from analogous solutions in Ref. \cite{YangDSI} for the local DSI equations under the reverse variable transformations $x\to -ix$, $t\to -t$, $Q\to -Q$, and accounting for the sign switching of the nonlinearity parameter $\epsilon$. These fundamental rational solutions of the nonlocal DSII equations are given by the same formulae (\ref{PPTAxyt})-(\ref{abomega}), except that the expressions for parameters $b$ and $\Delta$ are different:
\[
b\equiv \frac{p-\epsilon/p}{2}, \quad \Delta\equiv \frac{|p|^2}{(p+p^*)^2}.
\]
As before, $p$ and $\theta$ are free complex constants. Analysis of these rational solutions shows that they become rogue waves when $\epsilon=-1$ and $|p|=1$, in which case $a, \omega$ are imaginary and $b$ real. These rogue waves approach a constant background as $t\to -\infty$, but develop finite-time singularity at time $t_c=\theta_2/\omega_2$ and on the hyperbola
\[
(b_1y+\theta_1)^2-a_2^2x^2+\Delta=0.
\]
Graphs of these rogue waves are qualitatively similar to those in Fig.~1.

Multi-rogue waves in the nonlocal DSII equations (\ref{PPTDS}) can be derived from those of the local DSI equations in \cite{Satsuma_Ablowitz,YangDSI} under variable transformations and parameter conditions of $\epsilon=-1$, $|p_j|=1, j=1, \dots, n$. Details are omitted.

\section{Rogue waves in the nonlocal derivative NLS equation}
Now, we consider rogue waves in the nonlocal derivative NLS equation (\ref{e:PTDNLS}), i.e.,
\[ \label{e:PTDNLS2}
iu_t(x,t)+u_{xx}(x,t)+\sigma \left[u^2(x,t)u^*(-x,t)\right]_x=0,
\]
where $\sigma=\pm 1$.
These rogue waves can be obtained from rational solutions of the local derivative NLS equation (\ref{e:PTDNLSL}) through the variable transformation (\ref{transDNLS}). This local equation is invariant when $\sigma\to -\sigma, x\to -x$, thus we fix $\sigma=1$ without loss of generality. For this $\sigma$ value, the fundamental rational solution in the local derivative NLS equation is given by Eq.~(47) in Ref.~\cite{GuoDNLS}, and it is a moving soliton on a constant background. Then under the reverse of the transformation (\ref{transDNLS}), i.e., $x\to -ix$, $t\to -t$, this moving soliton of the local equation is converted to the following fundamental rational solution of the nonlocal equation,
\[ \label{e:DNLSf}
u(x,t)=\frac{(2ix-6t-i)(2ix-6t+3i)}{(2ix-6t+i)^2}.
\]
The graph of this solution is displayed in Fig.~2(a). It is seen that this is a rogue wave, rising from a constant background and then retreating back to the same background, analogous to the Peregrine solution of the NLS equation. However, the present rogue wave blows up to infinity at $x=-1/2$ and finite time $t_c=0$, unlike the Peregrine solution.

\begin{figure}[h!]
\centerline{\includegraphics[width=0.7\textwidth]{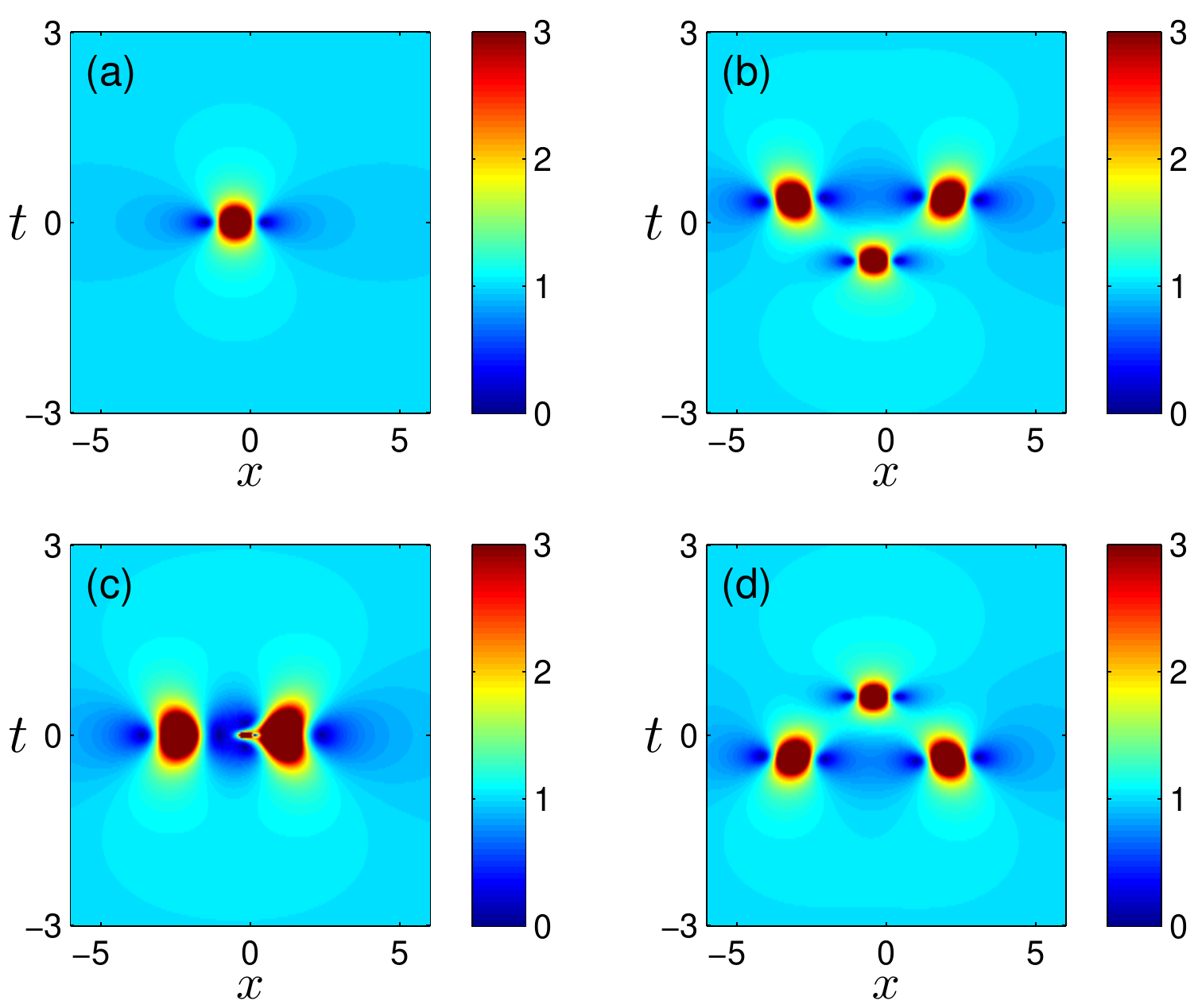}}
\caption{(color online) Rogue waves ($|u|$) in the nonlocal derivative NLS equation (\ref{e:PTDNLS2}) with $\sigma=1$. (a) The fundamental rogue wave (\ref{e:DNLSf});  (b, c, d) second-order rogue waves (\ref{e:DNLS2nd}) with $\alpha=-10, 0$ and $10$ respectively.}
\label{f:fig2}
\end{figure}

Higher-order rogue waves in the nonlocal equation (\ref{e:PTDNLS2}) can be obtained from higher-order rational solutions of the local derivative NLS equation. For instance, the second-order rational solution of the local equation is given by Eq.~(48) in Ref.~\cite{GuoDNLS}.
Then under the transformations $x\to -ix$, $t\to -t$, we get the second-order rational solution of the nonlocal equation (\ref{e:PTDNLS2}) as
\[ \label{e:DNLS2nd}
u(x,t)=\frac{(F_1-iF_2)F_3}{(F_1+iF_2)^2},
\]
where
\[
F_1=8\eta^3+18\eta-48t+12\alpha, \quad F_2=12\eta^2+3,    \nonumber
\]
\[
F_3=8\eta^3-30\eta-48t+12\alpha+i(36\eta^2-15),  \quad \eta=ix-3t,   \nonumber
\]
and $\alpha$ is a free real constant. Graphs of these rational solutions are plotted in Fig.~2(b,c,d) for $\alpha=-10, 0$ and $10$ respectively. It is seen that these rational solutions are second-order rogue waves which arise from and retreat back to the same constant background. But they blow up to infinity at three points of the $(x,t)$ plane.

\section{Multi-solitons and quasi-periodic solutions in the reverse space-time CMKdV equation}
In this section, we derive analytical solutions for the reverse space-time CMKdV equation (\ref{e:NCMKdV}), i.e.,
\[ \label{e:NCMKdV2}
q_{t}(x,t)+q_{xxx}(x,t)+6\sigma q(x,t)q^{*}(-x,-t)q_{x}(x,t)=0,
\]
where $\sigma=\pm 1$. The case of $\sigma=1$ will be called the focusing case, and that of $\sigma=-1$ the defocusing case. As we have shown, this nonlocal equation, under transformations (\ref{transCMKdV}), becomes the local CMKdV equation (\ref{e:CMKdV}) with the opposite sign of nonlinearity. Thus, we will derive analytical solutions for the defocusing/focusing nonlocal CMKdV equation from those of the local focusing/defocusing CMKdV equation.

\subsection{Multi-solitons in the nonlocal focusing equation}
Eq. (\ref{e:NCMKdV2}) in the focusing case has $\sigma=1$. Solitons and multi-solitons in this nonlocal focusing equation can be constructed from singular solutions in the local defocusing equation (\ref{e:CMKdV}). The local defocusing equation admits the following singular solutions
\begin{eqnarray}
q(x,t)=\sqrt{\nu} \exp(i \phi) \sec \hspace{0.04cm} [\sqrt{\nu} \hspace{0.06cm} (x+\nu t+\delta)], \quad \nu>0,
\end{eqnarray}
where $\nu$, $\phi$ and $\delta$ are real constants. Then under the reverse of transformations (\ref{transCMKdV}), i.e., $x\to -ix$, $t\to it$, this singular solution becomes
\[
q(x,t)=\sqrt{\nu} \exp(i \phi) \hspace{0.06cm} \mbox{sech} \hspace{0.04cm} [\sqrt{\nu} \hspace{0.06cm} (x- \nu t + i \delta)], \quad \nu>0,
\]
which is the fundamental soliton in the nonlocal focusing CMKdV equation (\ref{e:NCMKdV2}). Its peak amplitude, which occurs at $x-\nu t=0$, is $\sqrt{\nu}\sec(\sqrt{\nu}\hspace{0.05cm}\delta)$. Thus, for a given $\nu$, this peak amplitude can vary from $\sqrt{\nu}$ to infinity depending on the choice of the $\delta$ values.

Second-order singular solutions in the local defocusing equation (\ref{e:CMKdV}) can be obtained from Ref.~\cite{CMKdV2soliton} under certain parameter constraints [specifically, by requiring $c_1, c_2$ negative in Eq. (3.18) of that paper]. Then, under the above variable transformations, we get two-soliton solutions in the nonlocal focusing CMKdV equation (\ref{e:NCMKdV2}) as
\[ \label{e:CMKdV2soliton}
 q(x,t)=\frac{{\kappa} \left[ \sqrt{\nu_{1}}\exp(i\phi_{1})\cosh({\theta_{2}})+\sqrt{\nu_{2}}\exp(i\phi_{2})\cosh({\theta_{1}}) \right]}
 {({\kappa}^2-1)\cos(\phi_{1}-\phi_{2})+{\kappa}^2\cosh({\theta_{1}}-{\theta_{2}})+\cosh({\theta_{1}}+{\theta_{2}})},
\]
with
\[
{\kappa}=\frac{\sqrt{\nu_1}+\sqrt{\nu_2}}{\sqrt{\nu_1}-\sqrt{\nu_2}}, \quad {\theta_{k}}=\sqrt{\nu_{k}} \hspace{0.08cm} {\xi_{k}},  \nonumber
\]
\[
{\xi_{k}}=x-\nu_{k}t+i \hspace{0.04cm} \delta_{k},  \quad  k=1,2.   \nonumber
\]
For parameter choices
\[ \label{CMKdVpara}
\nu_{1}=1, \hspace{0.15cm} \nu_{2}=3, \hspace{0.15cm} \phi_{1}=\phi_{2}=0, \hspace{0.15cm}  \delta_{1}=\frac{1}{2}, \hspace{0.15cm} \delta_{2}=-\frac{1}{2},
\]
this two-soliton solution is displayed in Fig. 3(a).

\begin{figure}[h!]
\centerline{\includegraphics[width=0.7\textwidth]{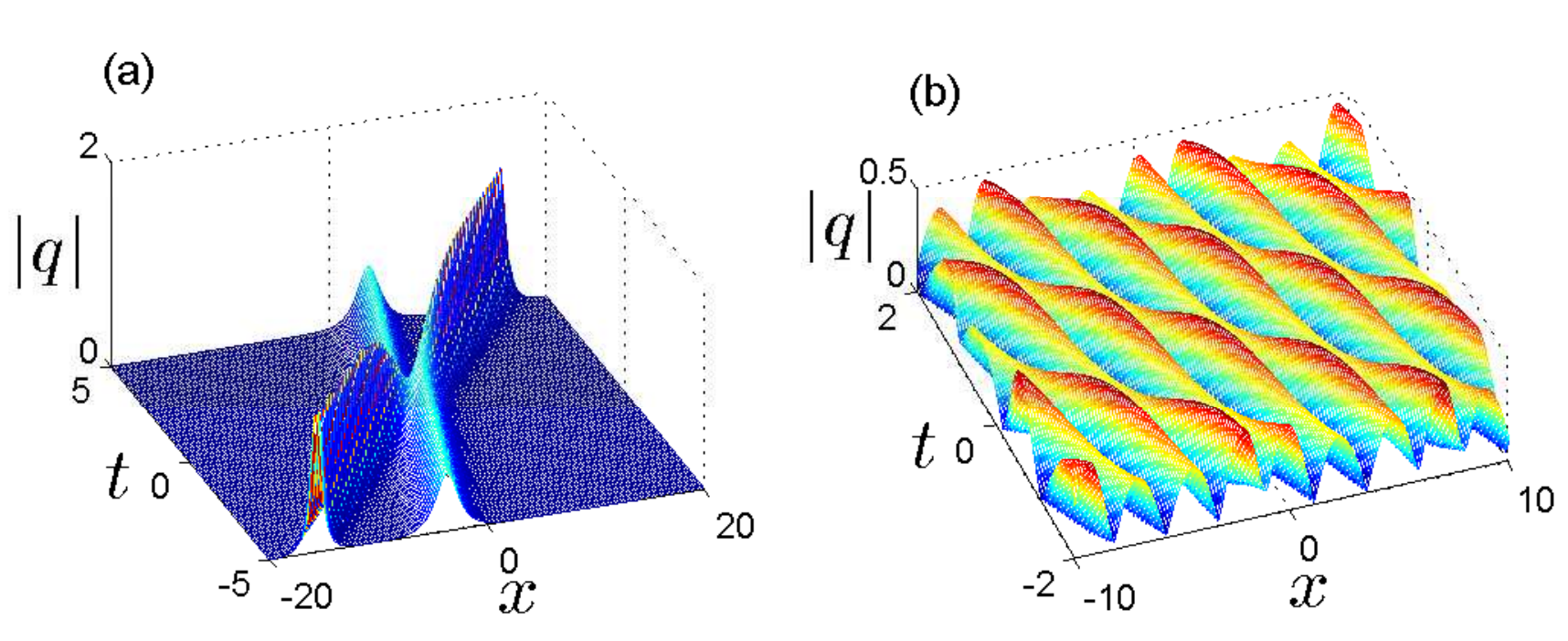}}
\caption{(color online) (a) A two-soliton solution (\ref{e:CMKdV2soliton}) in the nonlocal focusing CMKdV equation (\ref{e:NCMKdV2}) under parameter choices (\ref{CMKdVpara}). (b) A quasi-periodic solution (\ref{e:CMKdVquasi}) in the nonlocal defocusing CMKdV equation (\ref{e:NCMKdV2}) under parameter choices (\ref{CMKdVpara2}). }
\label{f:fig3}
\end{figure}

Higher-order solitons in the nonlocal focusing CMKdV equation (\ref{e:NCMKdV2}) can be obtained similarly.

\subsection{Quasi-periodic solutions in the nonlocal defocusing equation}
Eq. (\ref{e:NCMKdV2}) in the defocusing case has $\sigma=-1$. In this case, the local focusing CMKdV equation (\ref{e:NCMKdV}) has a soliton solution
\begin{eqnarray}
q(x,t)=\sqrt{c} \exp(i \phi) \hspace{0.04cm} \mbox{sech} \hspace{0.04cm} [\sqrt{c} \hspace{0.07cm} (x-c \hspace{0.04cm} t+\delta)], \quad c>0,
\end{eqnarray}
where $c, \phi$ and $\delta$ are real constants. Then under the transformations, we obtain the following solution for the nonlocal defocusing equation (\ref{e:NCMKdV2}),
\begin{eqnarray}
q(x,t)=\sqrt{c}\exp(i \phi) \sec \hspace{0.04cm} [\sqrt{c} \hspace{0.07cm} (x+c \hspace{0.04cm} t+i \hspace{0.04cm} \delta)], \quad c>0.
\end{eqnarray}
This solution is a traveling wave and is periodic in both space $x$ and time $t$.

Second-order extensions of this periodic solution can be obtained from two-soliton solutions of the local focusing CMKdV equation. The latter can be found in Eq. (3.18) of Ref. \cite{CMKdV2soliton}. Then, under variable transformations $x\to -ix$, $t\to it$, we get the following solution for the defocusing nonlocal CMKdV equation (\ref{e:NCMKdV2}) as
\[ \label{e:CMKdVquasi}
 q(x,t)=\frac{\kappa \left[\sqrt{c_{1}}\exp(i \hspace{0.04cm} \phi_{1})\cos(\theta_{2})+\sqrt{c_{2}}\exp(i \hspace{0.04cm} \phi_{2})\cos(\theta_{1})\right]}
 {(\kappa^2-1)\cos(\phi_{1}-\phi_{2})+\kappa^2\cos(\theta_{1}-\theta_{2})+\cos(\theta_{1}+\theta_{2})},
\]
where
\[
\kappa =\frac{\sqrt{c_1}+\sqrt{c_2}}{\sqrt{c_1}-\sqrt{c_2}}, \quad \theta_{j}=\sqrt{c_{j}} \hspace{0.08cm} \xi_{j}, \quad \xi_{j}=x+c_{j} \hspace{0.04cm} t+i \hspace{0.04cm} \delta_{j}.  \nonumber
\]
This solution contains two frequencies and is generically quasi-periodic in both space and time.
Under the parameter choices
\[ \label{CMKdVpara2}
c_{1}=1, \hspace{0.15cm} c_{2}=3, \hspace{0.15cm} \phi_{1}=\phi_{2}=0, \hspace{0.15cm}  \delta_{1}=\frac{1}{2}, \hspace{0.15cm} \delta_{2}=-\frac{1}{2},
\]
this double-frequency quasi-periodic solution is displayed in Fig. 3(b).

Higher-order quasi-periodic solutions in the nonlocal defocusing CMKdV equation (\ref{e:NCMKdV2}) can be obtained from higher-order solitons of the local focusing CMKdV equation in a similar way.

\section{New nonlocal integrable equations} \label{sec:newnonlocal}
In this last section, we show how these variable transformations can be used to derive new integrable nonlocal equations from their local counterparts.

\subsection{Nonlocal complex short pulse equations}
As the first example, we consider the integrable local complex short pulse (CSP) equation~\cite{Feng_shortpulse,Feng3},
\[ \label{e:CSP}
q_{xt} + q+\frac{1}{2}\sigma \left(|q|^2q_x\right)_x=0, \quad \sigma=\pm 1,
\]
where $q(x,t)$ is a complex function.
Under variable transformations $x \to -ix$, we get an integrable reverse space nonlocal CSP equation,
\[ \label{e:CSPspace}
q_{xt}(x,t)-iq(x,t)+\frac{1}{2}i\sigma [q(x,t)q_x(x,t)q^*(-x,t)]_x=0.
\]
Under a different variable transformation $t\to it$, we get an integrable reverse time nonlocal CSP equation,
\[ \label{e:CSPtime}
q_{xt}(x,t)+i q(x,t)+\frac{1}{2}i\sigma[q(x,t)q_x(x,t)q^*(x,-t)]_x=0.
\]
Under the combined transformations $x\to -ix, t\to it$, we get an integrable reverse space-time nonlocal CSP equation,
\[ \label{e:CSPspacetime}
q_{xt}(x,t)+ q(x,t)-\frac{1}{2}\sigma[q(x,t)q_x(x,t)q^*(-x,-t)]_x=0.
\]
Notice that this last nonlocal equation admits a reduction of $q(x,t)$ being real-valued. Under this reduction, we get an integrable reverse space-time real short-pulse equation
\[ \label{e:RSPspacetime}
q_{xt}(x,t)+ q(x,t)-\frac{1}{2}\sigma[q(x,t)q_x(x,t)q(-x,-t)]_x=0,
\]
where $q(x,t)$ is a real function.

Infinite numbers of conservation laws for these new nonlocal short-pulse equations can be inferred directly from those of local short-pulse equations through the corresponding variable transformations. For instance, the first two conserved quantities of the local CSP equation (\ref{e:CSP}) are
\begin{eqnarray}
I_1&=&\int_{-\infty}^\infty \sqrt{1+\sigma \hspace{0.05cm} q_x(x,t) \hspace{0.05cm} q^*_x(x,t)} \hspace{0.1cm} dx, \\
I_2&=&\int_{-\infty}^\infty \frac{q_{xx}(x,t)}{q_x(x,t)\sqrt{1+\sigma \hspace{0.05cm} q_x(x,t) \hspace{0.05cm} q^*_x(x,t)}} \hspace{0.1cm} dx.
\end{eqnarray}
The former quantity has been reported in \cite{Feng_shortpulse,Feng3}, and we found the latter quantity by inspiration of conserved quantities for the Wadati-Konno-Ichikawa hierarchy (which contains the real short pulse equation) \cite{Franca2012}. Then, using these conserved quantities and the transformation $x \to -ix$, we obtain the first two conserved quantities of the reverse space nonlocal CSP equation (\ref{e:CSPspace}) as
\begin{eqnarray}
I_1&=&\int_{-\infty}^\infty \sqrt{1-\sigma \hspace{0.05cm} q_x(x,t) \hspace{0.05cm} q^*_x(-x,t)} \hspace{0.1cm} dx, \\
I_2&=&\int_{-\infty}^\infty \frac{q_{xx}(x,t)}{q_x(x,t)\sqrt{1-\sigma \hspace{0.05cm} q_x(x,t) \hspace{0.05cm} q^*_x(-x,t)}} \hspace{0.1cm} dx.
\end{eqnarray}
Under the transformation $t\to it$, we obtain the first two conserved quantities of the reverse time nonlocal CSP equation (\ref{e:CSPtime}) as
\begin{eqnarray}
I_1&=&\int_{-\infty}^\infty \sqrt{1+\sigma \hspace{0.05cm} q_x(x,t) \hspace{0.05cm} q^*_x(x,-t)} \hspace{0.1cm} dx, \\
I_2&=&\int_{-\infty}^\infty \frac{q_{xx}(x,t)}{q_x(x,t)\sqrt{1+\sigma \hspace{0.05cm} q_x(x,t) \hspace{0.05cm} q^*_x(x,-t)}} \hspace{0.1cm} dx.
\end{eqnarray}
Under the combined transformations $x\to -ix, t\to it$, we obtain the first two conserved quantities of the reverse space-time nonlocal CSP equation (\ref{e:CSPspacetime}) as
\begin{eqnarray}
I_1&=&\int_{-\infty}^\infty \sqrt{1-\sigma \hspace{0.05cm} q_x(x,t) \hspace{0.05cm} q^*_x(-x,-t)} \hspace{0.1cm} dx,  \label{e:CSPST1} \\
I_2&=&\int_{-\infty}^\infty \frac{q_{xx}(x,t)}{q_x(x,t)\sqrt{1-\sigma \hspace{0.05cm} q_x(x,t) \hspace{0.05cm} q^*_x(-x,-t)}} \hspace{0.1cm} dx. \label{e:CSPST2}
\end{eqnarray}
The first two conserved quantities of the reverse space-time real short-pulse equation (\ref{e:RSPspacetime}) are simply these $I_1, I_2$ in
(\ref{e:CSPST1})-(\ref{e:CSPST2}) with the complex conjugate removed.

Higher conserved quantities of these new nonlocal CSP equations can be similarly obtained.

\subsection{Nonlocal nonlinear diffusion equations}
As a second example, we consider the local integrable NLS equation
\[ \label{e:NLS6}
iu_t+u_{xx}+2\sigma |u|^2u=0, \quad \sigma=\pm 1.
\]
Under the variable transformation $t\to -it$, we get an integrable reverse time nonlinear diffusion equation
\[ \label{e:nonlocal_diffision1}
u_t(x,t)-u_{xx}(x,t)-2\sigma u^2(x,t)u^*(x,-t)=0.
\]
Under the variable transformations $x\to ix, t\to it$, we get an integrable reverse space-time nonlinear diffusion equation
\[ \label{e:nonlocal_diffision2}
u_t(x,t)-u_{xx}(x,t)+2\sigma u^2(x,t)u^*(-x,-t)=0.
\]
Notice that both of these nonlocal equations admit the reduction of $u(x,t)$ being real. Under this reduction, we also obtain integrable reverse-time and reverse-space-time real nonlinear diffusion equations
\[ \label{e:realdiffusion1}
u_t(x,t)-u_{xx}(x,t)-2\sigma u^2(x,t)u(x,-t)=0,
\]
and
\[ \label{e:realdiffusion2}
u_t(x,t)-u_{xx}(x,t)+2\sigma u^2(x,t)u(-x,-t)=0.
\]

Infinite numbers of conservation laws for these new nonlocal diffusion equations can be readily derived from those of the local NLS equation (\ref{e:NLS6}) through variable transformations. For instance, using the first four conserved quantities of the local NLS equation~\cite{Yang2010}, we immediately obtain the first four conserved quantities of the reverse time nonlinear diffusion equation (\ref{e:nonlocal_diffision1}) as
\begin{eqnarray*}
&& \hspace{-1cm} I_1=\int_{-\infty}^\infty u(x,t)u^*(x, -t)dx, \\
&& \hspace{-1cm} I_2=\int_{-\infty}^\infty u^*(x, -t)u_x(x,t)dx, \\
&& \hspace{-1cm} I_3=\int_{-\infty}^\infty \left[u^*(x, -t)u_{xx}(x,t)+u^2(x,t)u^{*2}(x, -t)\right] dx, \\
&& \hspace{-1cm} I_4=\int_{-\infty}^\infty u^*(x, -t)\left\{ u_{xxx}(x,t)+[u^2(x,t)u^*(x, -t)]_x   +2u^*(x,-t)u(x,t)u_x(x,t)\right\}dx.
\end{eqnarray*}
Likewise, the first four conserved quantities of the reverse space-time nonlinear diffusion equation (\ref{e:nonlocal_diffision2}) are found to be
\begin{eqnarray*}
&& \hspace{-1cm} I_1=\int_{-\infty}^\infty u(x,t)u^*(-x, -t)dx, \\
&& \hspace{-1cm} I_2=\int_{-\infty}^\infty u^*(-x, -t)u_x(x,t)dx, \\
&& \hspace{-1cm} I_3=\int_{-\infty}^\infty \left[-u^*(-x, -t)u_{xx}(x,t)+u^2(x,t)u^{*2}(-x, -t)\right] dx, \\
&& \hspace{-1cm} I_4=\int_{-\infty}^\infty u^*(-x, -t)\left\{-u_{xxx}(x,t)+[u^2(x,t)u^*(-x, -t)]_x  +2u^*(-x,-t)u(x,t)u_x(x,t)\right\}dx.
\end{eqnarray*}
Conserved quantities for the reverse-time and reverse-space-time real nonlinear diffusion equations (\ref{e:realdiffusion1})-(\ref{e:realdiffusion2}) are simply those of the complex equations above with the conjugation removed.

Analytical solutions to these nonlocal diffusion equations can also be derived from solutions of the local NLS equation through transformations. For example, from the soliton solution of the local focusing NLS equation (\ref{e:NLS6}),
\[
u(x,t)=\eta \hspace{0.05cm} \mbox{sech} \hspace{0.05cm} [\eta (x-c \hspace{0.04cm} t)]\exp\left\{\frac{1}{2}i \hspace{0.04cm} c \hspace{0.04cm} x+i\left(\eta^2-\frac{1}{4}c^2\right)t\right\},
\]
with $\eta, c$ being real constants, we obtain the solution to the reverse time nonlocal diffusion equation (\ref{e:nonlocal_diffision1}) with $\sigma=1$ as
\[ \label{e:diffusion1}
u(x,t)=\eta \hspace{0.05cm} \mbox{sech} \hspace{0.05cm} [\eta (x+i\hspace{0.04cm} c \hspace{0.04cm} t)]\exp\left\{\frac{1}{2}i \hspace{0.04cm} c \hspace{0.04cm}  x+\left(\eta^2-\frac{1}{4}c^2\right)t\right\}.
\]
This solution exponentially grows or decays depending on the sign of $\eta^2-c^2/4$. In addition, it periodically collapses at location $x=0$.
For $\eta=c=1$, this solution is illustrated in Fig.~4 (left panel).

\begin{figure}[h!]
\centerline{\includegraphics[width=0.7\textwidth]{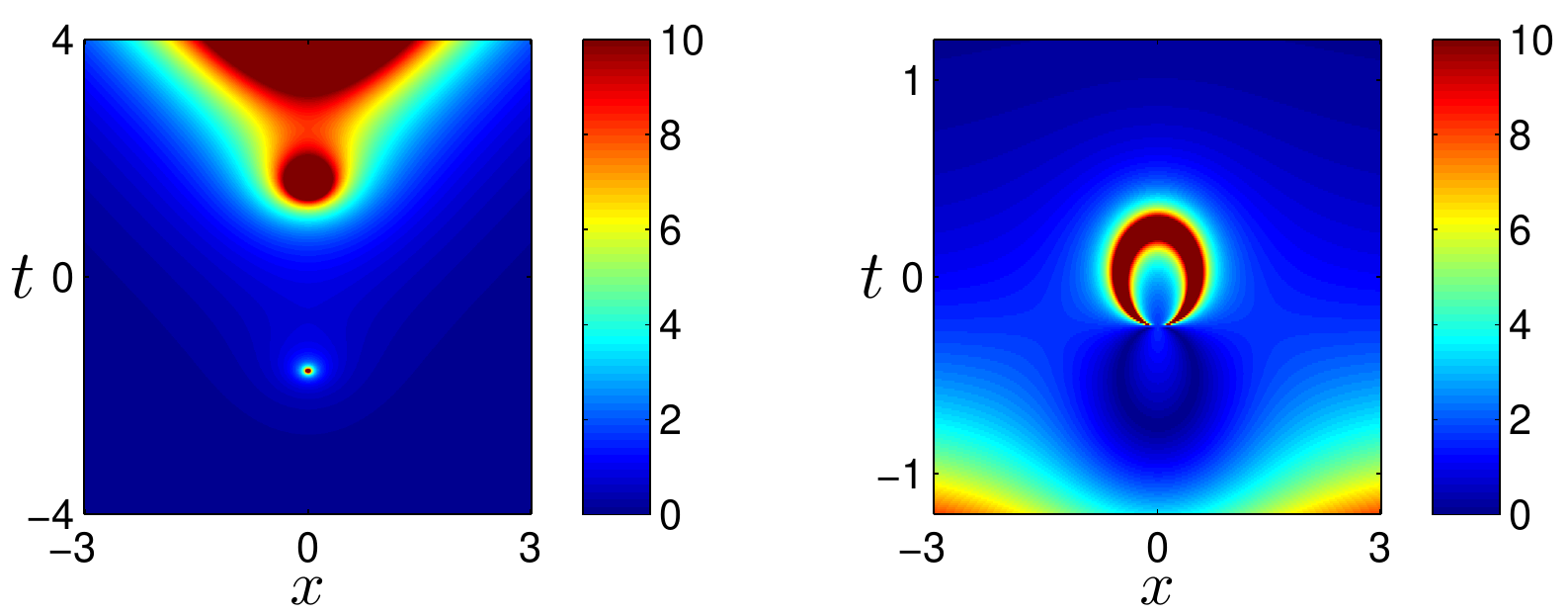}}
\caption{(color online) Left: solution (\ref{e:diffusion1}) to the reverse time nonlocal diffusion equation (\ref{e:nonlocal_diffision1}) with $\sigma=\eta=c=1$. Right: solution (\ref{e:diffusion2}) to the reverse space-time nonlocal diffusion equation (\ref{e:nonlocal_diffision2}) with $\sigma=1$. }
\label{f:fig4}
\end{figure}

As another example, from the Peregrine rogue wave solution of the local focusing NLS equation (\ref{e:NLS6}),
\[
u(x, t)=e^{2it}\left(1-\frac{4(1-4it)}{1+4x^2+16t^2}\right),
\]
and utilizing the transformations $x\to ix, t\to it$, we obtain the following solution to the reverse space-time nonlocal diffusion equation (\ref{e:nonlocal_diffision2}) with $\sigma=1$ as
\[ \label{e:diffusion2}
u(x, t)=e^{-2t}\left(1-\frac{4(1+4t)}{1-4x^2-16t^2}\right).
\]
This solution is illustrated in Fig.~4 (right panel). It decays exponentially with time, but blows up to infinity on the ellipse $4x^2+16t^2=1$ of the $(x,t)$ plane.

\subsection{Other new nonlocal integrable equations}
In addition to the above new nonlocal integrable equations, we can also obtain many other such equations using transformations. For instance,
from the local Sasa-Satsuma equation \cite{SS}
\[
u_t+u_{xxx}+6|u|^2u_x+3u\left(|u|^2\right)_x=0,
\]
and employing the variable transformations $x\to ix, t\to -it$, we get an integrable reverse space-time Sasa-Satsuma equation,
\begin{eqnarray}
&& \hspace{-1cm} u_t(x,t)+u_{xxx}(x,t)-6u(x,t)u^*(-x,-t)u_x(x,t)-3u(x,t)\left[u(x,t)u^*(-x,-t)\right]_x=0.
\end{eqnarray}
From the local Chen-Lee-Liu equation \cite{ChenLeeLiu}
\[
iu_t+u_{xx}+i|u|^2u_x=0,
\]
and employing the variable transformation $x\to ix, t \to -t$, we obtain an integrable reverse-space Chen-Lee-Liu equation
\[
iu_t(x,t)+u_{xx}(x,t)-u(x,t)u^*(-x,t)u_x(x,t)=0.
\]
A different transformation $x\to ix, t\to it$ yields a different integrable reverse-space-time nonlinear diffusion equation
\[
u_t(x,t)-u_{xx}(x,t)+u(x,t)u^*(-x,-t)u_x(x,t)=0.
\]
From the local modified NLS equation \cite{WadatiMNLS}
\[
iu_t+u_{xx}+i\alpha (|u|^2u)_x+\beta |u|^2u=0
\]
with real constants $\alpha, \beta$, and under transformations $x\to ix, t \to -t$, we get an integrable reverse-space modified NLS equation
\[
iu_t(x,t)+u_{xx}(x,t)-\alpha \left[u^2(x,t)u^*(-x,t)\right]_x-\beta u^2(x,t)u^*(-x,t)=0.
\]
From an integrable (2+1)-dimensional NLS equation \cite{2DNLS_lump_rogue}
\[
iq_t+q_{xy}+2iq(qq_x^*-q^*q_x)=0,
\]
and taking the transformations $x\to ix, y\to -iy$, we obtain an integrable reverse-space (2+1)-dimensional NLS equation
\begin{eqnarray}
&& \hspace{-1.3cm} iq_t(x,y,t)+q_{xy}(x,y,t)+2q(x,y,t)\left[q(x,y,t)q_x^*(-x,-y,t) -q^*(-x,-y,t)q_x(x,y,t)\right]=0.
\end{eqnarray}
Thus, this transformation technique is a powerful tool to generate a large class of new nonlocal integrable equations. Solution dynamics in these new nonlocal equations can also be studied through this transformation, as we have demonstrated earlier in this article.

\section{Summary}
In summary, we have reported that many recently proposed nonlocal integrable equations can be converted to local integrable equations through simple variable transformations. Examples include \PT-symmetric NLS and Davey-Stewartson equations, a nonlocal derivative NLS equation, the reverse space-time complex modified Korteweg-de Vries equation, reverse space-time three wave interaction equations, and many others. These transformations not only establish the integrability of these nonlocal equations, but also allow us to construct their analytical solutions from solutions of the local equations. These transformations can also be used to derive new nonlocal integrable equations. As applications of these transformations, we have used them to derive rogue wave solutions for the partially \PT-symmetric Davey-Stewartson equations and the nonlocal derivative NLS equation. In addition, we have used them to derive multi-soliton and quasi-periodic solutions in the reverse space-time complex modified KdV equation. Furthermore, we have used them to construct many new nonlocal integrable equations such as nonlocal short pulse equations, nonlocal nonlinear diffusion equations, nonlocal Sasa-Satsuma equations and nonlocal Chen-Lee-Liu equations.

These transformations reveal an intimate and deep connection between many nonlocal and local integrable equations. They are expected to provide a new and powerful tool in the study of these nonlocal equations.

\section*{Acknowledgment}
This material is based upon work supported by the Air Force
Office of Scientific Research under award number FA9550-12-1-0244, and the National Science Foundation under award
number DMS-1616122. The work of B.Y. is supported by a visiting-student
scholarship from the Chinese Scholarship Council.

\end{document}